\documentclass[11pt]{article}
\usepackage{cite}
\usepackage[usenames,dvipsnames]{pstricks}
\usepackage[utf8]{inputenc}
\pdfoutput=1
\usepackage[fleqn]{amsmath}
\usepackage{amsmath,amsfonts}
\usepackage[english]{babel}
\usepackage{textcomp}
\usepackage{caption,subcaption}
\usepackage{graphicx}
\usepackage{rotating}
\usepackage{authblk}
\usepackage{epstopdf}
\usepackage{bookmark}
\usepackage{float}
\usepackage{amsmath}
\usepackage[square,numbers]{natbib}
\usepackage{hypernat}
\usepackage{color, soul}
\hypersetup{pdfauthor={some author},pdftitle={eye-catching title}}
\usepackage{ifpdf}

\title{\textbf{Effects of electromagnetic field on a radiating star}}
\author[1]{Suresh C. Jaryal\footnote{suresh.fifthd@gmail.com}}
\author[2]{Ayan Chatterjee\footnote{ayan.theory@gmail.com}}
\author[2]{\small{Akshay Kumar \footnote{akshay.relativity@gmail.com}}}
\affil[1]{{Department of Physics \& Astronomical Sciences},
Central University of Jammu, Samba, J\&K- 181143, India.}
\affil[2]{Department of Physics \& Astronomical Science, Central University of Himachal Pradesh, Dharamshala- 176215, India.}
\numberwithin{equation}{section}
\begin{document}
\maketitle
In this paper we study the shear free spherical symmetric gravitational collapse of charged radiating star. All the physical quantities including pressure, density are regular. Energy conditions are satisfied throughout the interior of the matter configuration. The luminosity is time independent and mass is radiated linearly. The causal and non causal temperature remains greater than that of the uncharged collapsing scenario. 
\section{Introduction}

The quest to understand the formation of stars, horizons, cosmic censorship conjecture, and spacetime singularities has led to a substantial amount of work on gravitational collapse during the last hundred years. Beginning with the pioneering work of Oppenheimer-Snyder and Dutt on homogeneous dust collapse, also known as the OSD model~\cite{OS,D}, our knowledge of gravitational collapse has been considerably extended through the inclusion of matter inhomogeneities, tangential and radial pressures~\cite{Hawking_ellis, pankaj_book}. Implications of these features towards the collapse process have led to the development 
of interesting phenomenology directed to detect whether the end result is a black hole or a horizonless singularity~\cite{pankaj_book,Santos1985,Santos1987,KST,Herrera1989,Chan1997,Chan1998,pankaj_malafarine_review}.
Further, a number of exact physical solutions have been introduced to study the stability and dynamics of stellar objects ~\cite{Reissner,Nordstrom,Santos,HPO,SP2016,MGR,PB2017,LHAPJP,Govender:2020gyc,Govender:2021gyv}. However, the study of spherical stellar systems is incomplete without considering the effects of charge on its time development, as it plays an important role throughout its lifespan. The effects of charge in the stellar system were started with the pioneer works of Reissner–Nordström~\cite{Reissner,Nordstrom} and Bonnor~\cite{Bonnor1960,Bonnor1965}. 
It has been observed that the collapsing process is delayed due to the presence of the charge in the stellar system~\cite{DiPrisco:2007vgp,Guha:2013voa,YN_SDM,GZA_SDM,Maurya:2021jod,Nazar:2023aif,Maurya:2023xfo}. 
Indeed, the fact that the presence of electric/ magnetic charges are crucial in the time development of stellar as well as supermassive black holes at the center of galaxies, has been stressed several times~\cite{Mashhoon:1979tt,Zakharov:2005ek,Chrusciel:2012jk,Zakharov:2014lqa,Zajacek:2018ycb,Zajacek:2019kla,Gong:2019aqa}. However, several facets of such solutions, like thermodynamic stability, apparent luminosity, effect of temperature profiles and energy conditions on these physically acceptable solutions remains unknown. Furthermore, such solutions are context dependent, in the sense that the mathematical techniques are dependent on the functional forms of the choices of density, pressure and metric variables. In this paper, we show that, using the Karmarkar conditions, it is possible to understand the collapse process of a dissipative, shear- free collapse of a charged matter field admitting anisotropic stress, under generic choices of the matter and geometric variables. Indeed, the present method adapts itself naturally to anisotropy of matter fields and the dissipative
nature of the energy- momentum tensor.  We also show the precise dependence of these variables on the stability, luminosity and adiabatic index.

Before describing our model, let us understand the precise nature of difficulties arising in the study of dissipative stellar collapse models. First, note that the radiating gravitational collapse of a stellar object is an extreme dissipative process. This requires terms relating heat flux to geometric variables be introduced which are continuous and differentiable. Secondly, the continuous and smooth matching of the interior and exterior spacetimes over the timelike hypersurface $\Sigma$ requires that radial pressure to be proportional to the heat flux at the boundary of collapsing radiant star. Since a physically consistent charged gravitationally collapsing model must satisfy both Einstein- Maxwell field equations as well as the set of energy conditions, the nonlinearity of the Einstein- Maxwell field equations makes it difficult to find its exact solutions. In this paper, we show that these complications described above may be taken care of by the Karmarkar conditions~\cite{Eisenhart1925a,Eisenhart1925b,Karmarkar, Banerjee,NGM2018,ON}. For a shear free stellar system, the Karmarkar condition relates the two metric variables in terms of each other, and with a single choice of metric variables, the system is exactly solvable. We must however stress that our solution only represents astrophysical systems which are spherical or nearly spherical/ slowly rotating. Exact solution of non spherical collapse processes remains a major unsolved problem.

In this paper, we extend this Karmarkar formalism for a charged radiating collapsing stellar model. 
To determine the metric variables which satisfy the solutions of nonlinear field equations and energy conditions, we consider an adhoc but physical form of the pressure anisotropy. There  are two advantages of considering this particular form of pressure anisotropy.
First is using this form together with the Karmarkar condition make our model to be exactly solvable. Second, we do not have to consider an adhoc form of charged density, which was the requirement in the earlier studies of the charged stellar systems~\cite{Jaryal:2020mag,Jaryal:2021lsu}. We also show that our model is physical, as all the physical quantities are regular and also satisfy the standard energy conditions. Furthermore, our study of the thermodynamical evolution shows that the causal and non causal temperature for the charged case remains greater than that of the uncharged collapsing scenario, as required.

The present article has been organized as follows. In the following section \ref{sec2} we describe the shear free interior spacetime with composite matter distribution and the Einstein- Maxwell field equations. The interior shear free spaceime is smoothly matched with the exterior Reissner–Nordström spacetime across the timelike surface. In the section \ref{sec3}, we present a new class of exact solutions to the Einstein- Maxwell field equations and derive the explicit expressions for physical quantities. It is then shown that the said model is physically viable as all the physical quantities are regular. The study of the stability of stellar system, which shows that the system is unstable and the energy conditions are well satisfied throughout the collapse are given in section \ref{sec4}. In section \ref{sec5}, we give the thermodynamical evolution of the charged radiative gravitational collapse. Finally, discussion of the result accompanied with concluding remarks are in section \ref{sec6}.

\section{Matter distribution and the governing equations}\label{sec2}
In comoving coordinates, the general shear free spherically symmetric metric is given by 
\begin{eqnarray}
 ds^{2}=-a(r)^{2} dt^{2}+ b(r)^{2}\,f(t)^{2} dr^{2} + r^{2}\,b(r)^{2}\,f(t)^{2}\left(d \theta^{2}+ \sin^{2}{\theta} d\phi^{2} \right)
\label{1eq1}
\end{eqnarray}
Also, we consider the energy-momentum tensor for the charged anisotropic distribution 
with radial heat flow as
\begin{eqnarray}
T_{\mu\nu}&=&(p_{t}+\rho)u_{\mu} u_{\nu} +p_{t} g_{\mu\nu}+(p_{r}-p_{t})X_{\mu}X_{\nu}\nonumber\\
&&+q_{\mu}u_{\nu}+q_{\nu}u_{\mu}+\left( F^{\lambda}_{\mu}\,F_{\nu\lambda}
-\frac{1}{4}\,g_{\mu\nu} F^{\lambda\delta}\,F_{\lambda\delta}\right) \label{tmnz}
\end{eqnarray}
where $\rho$, $p_{r}$ and $p_{t}$ are the density, radial pressure and tangential pressure respectively,
$q^{\mu}$, $u^{\mu}$ and $X^{\mu}$ are  and radial heat flow vector, unit time-like $4$-velocity vector and unit space-like vector along radial $4$-vector respectively, satisfying $u_\mu u^\mu=-X_\mu X^\mu=-1$ and $u_\mu X^\mu=u_\mu q^\mu=0$. And $F_{\mu\nu}$ represent the Maxwell field tensor.
In the comoving coordinates, the radial heat flow vector $q^{\mu}$, $4$-velocity $u^{\mu}$ and unit space-like vector $X^{\mu}$ of the fluid are given by 
 \begin{eqnarray}
  \hspace{2cm}q^\mu=\frac{1}{b\,f}\,X^{\mu}\hspace{0.5cm} ;\hspace{0.5cm} u^\mu=\frac{1}{a}\,\delta^{\mu}_0 \hspace{0.5cm};\hspace{0.5cm} X^\mu=\frac{1}{b\,f}\,\delta^{\mu}_1   \label{uXq}.
 \end{eqnarray}
The expansion scalar, $\Theta$ for the metric \eqref{1eq1} is given by
\begin{eqnarray}
\Theta&=&\bigtriangledown_{\mu}u^{\nu}=\frac{3\,\dot{f}}{a\,f}\label{Theta}.
\end{eqnarray}
 The Maxwell's equations are given by
 \begin{eqnarray}
 F_{\mu\,\nu}&=&\phi_{\nu,\,\mu}-\phi_{\mu,\,\nu} \label{Fab}\\
 F^{\mu\,\nu}\,_{;\nu}&=&J^{\mu} \label{Fa;b}
 \end{eqnarray}
 where $\phi_{\mu}$ and $J^{\mu}$ represents the four potential and four current respectively.
 It must be noted that the charge is assumed to be at rest with respect to the 
 comoving coordinates of the metric \eqref{1eq1}, which implies that we have no magnetic field present,
 thus the four potential $\phi_{\mu}$ and four current $J^{\nu}$ takes the form as
\begin{eqnarray}
\phi_{\mu}&=&\phi(t,r)\,\delta^{0}_{\mu}\hspace{1cm} ;\hspace{1cm} J^{\mu}=\sigma(t,r) u^{\mu} \label{phiJ}
\end{eqnarray}
Using the choice of the potential from equation \eqref{phiJ} into the equation \eqref{Fab},
the nonzero component of $F_{\mu\nu}$ are
\begin{eqnarray}
F_{01}&=&-F_{10}=-\frac{\partial \phi}{\partial r} \label{F01}
\end{eqnarray}
Now, for the interior spacetime \eqref{1eq1} and \eqref{uXq}, using equations \eqref{phiJ} 
and \eqref{F01} into the Maxwell's  equations \eqref{Fab} and \eqref{Fa;b} we have
\begin{eqnarray}
\phi^{''}-\left(\frac{b^{'}}{b}-\frac{a^{'}}{a}+\frac{2}{r} \right)\phi^{'}&=&ab^2f^2\sigma\label{M1}\\
\dot{\phi}\,^{'}+\frac{\dot{f}}{f}\phi^{'}=0 \label{M2}
\end{eqnarray}
On integration of equation \eqref{M1} we obtained
\begin{eqnarray}
\phi^{'}&=&\frac{s\,a}{r^2\,b\,f}\label{phi'}
\end{eqnarray}
the law of conservation of charge implies that 
\begin{eqnarray}
s(r)&=&\int_{0}^{r} r^2\,\sigma\,b^3\,f^3\,dr \label{s(r)}
\end{eqnarray}
where $s(r)$ is the time independent total electric charge $Q$ within a radius $r$.
The non vanishing components of Einstein-Maxwell field equations for the 
 metric \eqref{1eq1}, energy momentum tensor \eqref{tmnz} and  \eqref{uXq}  are $($using units with $c=1=8\pi G)$ 
 \begin{eqnarray}
\rho+\frac{1}{2}\frac{s^2}{r^4\,b^4\,f^4}&=&\frac{3\,\dot{f}^{2}}{a^2\,f^2}
-\frac{1}{b ^{2}\,f^{2}}\left(\frac{2\,b^{''}}{b}-\frac{b^{'}\,^{2}}{b\,^{2}}+\frac{4}{r}\frac{b^{'}}{b} \right)\label{rho},\\
{p}_{r}-\frac{1}{2}\frac{s^2}{r^4\,b^4\,f^4}&=&-\frac{1}{a^{2}}\left(\frac{2\,\ddot{f}}{f}
+\frac{\dot{f}^{2}}{f^{2}}\right)+\frac{1}{b^{2}\,f^{2}}\left(\frac{2\,a^{'}\,b^{'}}{a\,b}
+\frac{2}{r}\left(\frac{a^{'}}{a}+\frac{b^{'}}{b}\right)+\frac{b^{'}\,^{2}}{b\,^{2}} \right)
\label{pr},\\
 {p}_{t}+\frac{1}{2}\frac{s^2}{r^4\,b^4\,f^4}&=&-\frac{1}{a^{2}}\left(\frac{2\,\ddot{f}}{f}
 +\frac{\dot{f}^{2}}{f^{2}}\right)+\frac{1}{b^{2}\,f^{2}}\left(\frac{a^{''}}{a}
 +\frac{1}{r}\left(\frac{a^{'}}{a}+\frac{b^{'}}{b}\right)-\frac{b^{'}\,^{2}}{b\,^{2}} +\frac{b^{''}}{b}\right)\label{pt},\\
 q&=&-\frac{2\,a^{'}\,\dot{f}}{a^{2}\,b^{2}\,f^{3}}\label{q}.
 \end{eqnarray}
where dot and prime are the derivatives with respect to $t$ and $r$ respectively.
From equations \eqref{rho}-\eqref{q} we can see that the number of field equations 
are less than that of the number of unknowns. Also, the form of the three metric 
potentials fixes all the unknown physical quantities of the system. So, in order to completely study
the collapsing phenomena we need to find the forms of these metric potentials.

First, let us give the Israel-Darmois  junction conditions. The interior manifold is $\it\bf{M^-}$
and exterior manifold is $\it\bf{M^+}$ to be matched across the bounding timelike three space $\Sigma$, at  $r=r_b$. 
The exterior spacetime is described by the 
Vaidya– Reissner– Nordstrm spacetime having outgoing radial flow of the radiation  around a charged spherically symmetric
source of gravitational field is given by\,\cite{Reissner,Nordstrom}
\begin{eqnarray}
ds^{2}_{+}=-\left(1-\frac{2M(v)}{{\bf{r}}}+\frac{Q^2}{\bf{r}^2}\right)dv-2dvd{\bf{r}}+ {\bf{r}}^2 \left(d \theta^2
 +\sin^2\theta d\phi^2\right)\label{m+}
\end{eqnarray}
where $M(v)$ and $Q$ are the total mass and total charge respectively.
The junction conditions require the matching of metric as well as the extrinsic curvatures
\begin{eqnarray}
ds^2_{\Sigma}&=&(ds_{-})^2_\Sigma = (ds_{+})^2_\Sigma \label{ds}\\
\left[K_{ij}\right]_{\Sigma}&=&K_{ij}^{+}=K_{ij}^{-} \label{Kij}
\end{eqnarray}
where $K_{ij}^{\pm}$ is the extrinsic curvature to $\Sigma$. 
The junction condition \eqref{ds} at the hypersurface $\Sigma$ gives
\begin{eqnarray}
dt&=&a(r)_{_{\Sigma}}^{-1}\, d\tau , \label{ttau}\\
{\bf{r}}_{_{\Sigma}}(v)&=&(r\,b\,f)_{_{\Sigma}} ,\label{rR}\\ 
\left(\frac{dv}{d\tau}\right)_{\Sigma}^{-2}&=&\left( 1-\frac{2M}{\bf{r}}+\frac{Q^2}{\bf{r}^2}+2\frac{d\bf{r}}{dv}\right)_{\Sigma}. \label{vtau}
\end{eqnarray}
where $\tau$ is the time coordinate defined only on the hypersurface $\Sigma$.
The unit normal vectors on the hypersurface $\Sigma$ for the interior and exterior spacetime are given by
\begin{eqnarray}
n^{-}_{l}&=&\left[0,(b\,f)_{_{\Sigma}},0,0\right]\label{n-}\\
n^{+}_l&=&\left(1-\frac{2M(v)}{{\bf{r}}}+\frac{Q^2}{\bf{r}^2}+2\frac{d{\bf{r}}}{dv}\right)^{-\frac{1}{2}}_\Sigma 
\left(-\frac{d{\bf{r}}}{dv}\delta^0_l+\delta^1_l\right)_{\Sigma} \label{n+}
\end{eqnarray}
The non vanishing components of the extrinsic curvature for metrics \eqref{1eq1} and \eqref{m+} are given by
\begin{eqnarray}
K^{-}_{\tau\tau}&=&-\left[\frac{a^{'}}{a\,b\,f}\right]_{\Sigma}, \label{Ktt-} ~~~
K^{-}_{\theta\theta}=\left[r\,b\,f\left(1+\frac{r\,b^{'}}{b} \right)\right]_\Sigma , \label{Kthth-}\\
K^{+}_{\tau\tau}&=& \left[\frac{d^2v}{d\tau^2}\left(\frac{dv}{d\tau}\right)^{-1}-\left(\frac{dv}{d\tau}\right)
\left(\frac{M}{{\bf{r}}^2}-\frac{Q^{2}}{{\bf{r}}^{3}}\right)\right]_\Sigma ,\label{Ktt+}\\
K^{+}_{\theta\theta}&=&\left[\left(\frac{dv}{d\tau}\right)\left(1-\frac{2M}{\bf{r}}+\frac{Q^2}{{\bf{r}}^2}\right)
{\bf{r}}-{\bf{r}}\frac{d{\bf{r}}}{d\tau}\right]_\Sigma ,\label{Kthth+}\\
K^{-}_{\phi\phi}&=& \sin^2{\theta} K^{-}_{\theta\theta}\,\, , \,\, K^{+}_{\phi\phi}=\sin^2{\theta}
 K^{+}_{\theta\theta} .\nonumber
\end{eqnarray}
Now, from the second junction condition \eqref{Kij}  $K^{+}_{\theta\theta}=K^{-}_{\theta\theta}$
 at hypersurface $\Sigma$, and using equations \eqref{ttau}, \eqref{rR} and \eqref{vtau} gives
\begin{eqnarray}
\left[r\,b\,f\left(1+\frac{r\,b^{'}}{b} \right)\right]_\Sigma&=&
\left[\left(\frac{dv}{d\tau}\right)\left(1-\frac{2M}{{\bf{r}}}+\frac{Q^2}{{\bf{r}}^2}\right)
{\bf{r}}-{\bf{r}}\frac{d{\bf{r}}}{d\tau}\right]_\Sigma\label{Kth+-}\\
m_{\Sigma}&=&\left[\frac{r^{3}\,\dot{f}^{2}\,b^{3}f}{2\,a^{2}}-\frac{r^{3}\,f\,b^{'}\,^{2}}{2\,b}
-r^2\,f\,b^{'}+\frac{Q^2}{2rbf}\right]_{\Sigma} \label{mass}
\end{eqnarray}
where $2M$ is the total energy entrapped inside the hypersurface $\Sigma$\,\cite{DiPrisco:2007vgp, Misner-Sharp, CM}.
Now, again from the matching condition \eqref{Kij}, the matching of the $K^{+}_{\tau\tau}=K^{-}_{\tau\tau}$ 
component together with the equations \eqref{ttau} we have
\begin{eqnarray}
-\left[\frac{a^{'}}{a\,b\,f}\right]_{\Sigma}&=&\left[\frac{d^2v}{d\tau^2}\left(\frac{dv}{d\tau}\right)^{-1}
-\left(\frac{dv}{d\tau}\right)\left(\frac{M}{{\bf{r}}^2}-\frac{Q^{2}}{{\bf{r}}^{3}}\right)\right]_\Sigma \label{Ktt+-}
\end{eqnarray}
Substituting equations \eqref{ttau}, \eqref{rR} and \eqref{mass} into the equation \eqref{Kth+-} we have 
\begin{eqnarray}
\left(\frac{dv}{d\tau}\right)_{\Sigma}&=&\left(1+\frac{r\,b^{'}}{b}+\frac{r\,b\,\dot{f}}{a}\right)_{\Sigma}^{-1} \label{dvdtau}
\end{eqnarray}
This is the gravitational red shift as viewed by an observer at rest at infinity.
Now, differentiating \eqref{dvdtau} with respect to the $\tau$ and using equations 
\eqref{mass} and \eqref{dvdtau}, we can write the equation \eqref{Ktt+-} 
and comparing with equations \eqref{pr} and \eqref{q} we have
%
\begin{eqnarray}
(p\,_{r})_{_{\Sigma}}&=&(q\,b\,f)_{_{\Sigma}}. \label{matching}
\end{eqnarray}
We also require the  total Luminosity for an observer at rest at infinity is given by\,\cite{Chan1997}
%
\begin{eqnarray}
L_{\infty}&=& -\left( \frac{dm}{dv}\right)_{\Sigma}=-\left[ \frac{dm}{dt}
\frac{dt}{d\tau}\left(\frac{dv}{d\tau}\right)^{-1}\right]_{\Sigma}.\label{LInf}
\end{eqnarray}
%
%
\section{Exact solutions of the field equations} \label{sec3}
%
All the physical quantities in the  Einstein- Maxwell field equations \eqref{rho}-\eqref{q} depends upon the forms of metric variables. In our previous work, we have developed an adhoc formalism to obtained the exact solutions of the field equations, and the model obtained is shown to be physically viable both for the GR \cite{Jaryal:2020mag} and $f(R)$ regime \cite{Jaryal:2021lsu}. We consider an adhoc form of the pressure anisotropy so that we can use its equation to obtained the solution of one of metric variable. In this work we will generalize it to the charged collapsing scenarios. We used the Karmarkar condition to find the other metric variable.  From equations \eqref{pr}, \eqref{pt}, the pressure anisotropy $\Delta=p_{t}-p_{r}$ has the form
\begin{eqnarray}
\Delta&=&\frac{1}{f^2\,b^2}\left[\frac{a^{''}}{a}+\frac{b^{''}}{b}
-\frac{1}{r}\left(\frac{a^{'}}{a}+\frac{b^{'}}{b}\right)-\frac{2\,a^{'}\,b^{'}}{a\,b}-\frac{2\,b^{'\,^{2}}}{b^2}
\right]-\frac{s^2}{r^4\,b^4\,f^4} \label{Delta}
\end{eqnarray}
To find the exact solutions of the metric variable, we consider the following adhoc form of the pressure anisotropy $\Delta$ as
\begin{eqnarray}
\Delta &=& \frac{\alpha}{f^2\,b^2}\left[\frac{a^{''}}{a}-\frac{a^{'}}{r\,a}-\frac{2\,a^{'}\,b^{'}}{a\,b}\right]-\frac{s^2}{r^4\,b^4\,f^4}\label{Delta1}
\end{eqnarray}
The above form of the anisotropy parameter \eqref{Delta1} reduces the total pressure anisotropy equation \eqref{Delta} into a differential equation of one variable and the solution of metric variable $b(r)$ is given by
\begin{eqnarray}
0&=&\frac{1}{f^2\,b^2}\left[\frac{b^{''}}{b}-\frac{b^{'}}{r\,b}-\frac{2\,b^{'\,^{2}}}{b^2}  \right]\nonumber\\
b(r)&=&-\frac{2}{C_3\,r^2+2\,C_4}\label{br}
\end{eqnarray}
where $C_3$ and $C_4$ are constant of integration. The solution obtained for the $b(r)$ is the same as that of obtained for our earlier work, however the difference is in the choice of the pressure anisotropy equation \eqref{Delta1}. This choice of the charged pressure anisotropy give us the freedom to have the time dependent solutions. 

It has been shown that an $(n+1)$-dimensional space $V^{n+1}$ can be
 embedded into a pseudo Euclidean space $E^{n+2}$ of dimension $(n+2)$ 
 if there exists a symmetric tensor $b_{\mu\nu}$ which satisfies the 
 following Gauss- Codazzi equations \citep{Eisenhart1925a}.
\begin{eqnarray}
R_{\mu\,\nu\,\lambda\,\delta}&=&2\,e\,b_{\mu[\lambda}\,b_{\delta]\nu}\\
0&=&b_{\mu\,[\nu;\,\lambda]}-\Gamma^{\sigma}_{[\nu\,\lambda]}\,
 b_{\mu\,\sigma}+\Gamma^{\sigma}_{\mu\,[\nu}\, b_{\lambda]\,\sigma}
\end{eqnarray}
where $e=\pm 1$ ($+$ or $-$, when the normal to the manifold is spacelike or timelike respectively) and 
$b_{\mu\,\nu}$ are the coefficient of the second differential form. 
The necessary and sufficient condition for any Riemannian space to 
be embedding class $\bf{I}$, is also known as the Karmarkar conditions \cite{Eisenhart1925b, Karmarkar}
\begin{eqnarray}
R_{rtrt}\,R_{\theta\phi\theta\phi}=R_{r\theta r\theta}\,R_{t\phi t\phi}
-R_{\theta rt\theta}\,R_{\phi rt\phi}\label{Karmarkar}
\end{eqnarray}
As shown in \cite{Jaryal:2020mag,Jaryal:2021lsu}, for the shear free spherically symmetric metric \eqref{1eq1}, the Karmarkar  class {\bf{I}} condition becomes 
\begin{eqnarray}
0&=&r\,b^{3}{\dot{f}}^{2}\,\left(rab\,{a{''}}-ab\,{a{'}}
+{rb\,a{'}^{2}}-2\,r a\,a'b'\right)
+r\,a^{2}b^{2}f\ddot{f}\left(b\,b{'}+2r{a{'}^{2}}
-rb\,{b{''}} \right)\nonumber\\
&+&ra^{3}a{'}b{''}\left(rb'+b a' \right) -ra^{3}a{''}\left(2ba'
+r b'{^{2}}\right)
+a^{3}a'b'\left(b+2r b'\right)
\label{Karmarkar1}
\end{eqnarray}
The Karmarkar condition equation \eqref{Karmarkar1} is a nonlinear differential in its temporal and radial behavior. A physical model of collapse must simultaneously satisfy both the conditions \eqref{matching} and \eqref{Karmarkar1}. One of the solution, which satisfies both these conditions, is to have a linear temporal dependence in the model as considered in\,\cite{Banerjee}
\begin{eqnarray}
f(t)&=&-C_{Z} \,t \label{f(t)}
\end{eqnarray}
where $C_Z>0$ for the collapsing phenomena. Using the forms of $b(r)$ and $f(t)$ in equation \eqref{Karmarkar1} give rise to the form of $a(r)$ as
\begin{eqnarray}
a(r)&=&\frac{1}{2\sqrt{2}\sqrt{C_3C_4}}\sqrt{C_4^{2}\left(C_1b(r)+4C_2C_3\right)^2-4C_Z^2} \label{a(r)}
\end{eqnarray}
where $C_1$ and $C_2$ are integration constants.
These set of solutions are similar to as found in \cite{Jaryal:2020mag}, and there it has been shown that 
combination of the Karmarkar condition with the pressure isotropy leads to only two set of exact solutions which are:
the Schwarzschild\,\cite{KS} or Kohler and Chao\,\cite{KC} solution.
We knows that the Kohler- Chao solution is only physical for unbound configuration such as cosmological model as the radial pressure vanishes at $r\rightarrow \infty$.
Also, in the static case, from karmarkar condition and pressure isotropy we must have $C_{1}=0$. Thus, the Schwarzschild like form of the metric variables are
\begin{eqnarray}
a(r)^{2}=\frac{4C_{2}^{2}C_{3}^{2}C_{4}^{2}-1}{2C_{3}C_{4}}\hspace{0.3cm};\hspace{1cm}b(r)^{2}=\frac{4}{\left(2\,C_4+C_3\,r^2\right)^{2}}\hspace{0.3cm};\hspace{1cm} f(t)^{2}=1.
\end{eqnarray}
Recently Ospino et.al.\,\cite{ON}, derived the Karmarkar scalar condition, and  used for the nonstatic system, where they found two class of solutions. One of their solution represent the earlier found solution of horizon free radiating collapse as found by the Naidu et. al. \cite{NGM2018}. However, the other solution is the completely new class of solution. The form of second class of solutions are
\begin{eqnarray}
A(t,r)&=&-\dot{\bar{b}}(t) \frac{\sqrt{C\left(-2+\bar{a}C\right)+\bar{a}r^{4}\left(-1+\bar{a}\,^2 C^2\right)+2r^2\left(-1-\bar{a}C+\bar{a}\,^{2} C^2 \right)}}{{\sqrt{2} (1+\bar{a} r^2)}}\label{A}\\
B(t,r)&=&\frac{\bar{b}(t)}{(1+\bar{a} r^2)}\hspace{0.5cm};\hspace{1cm} R(t,r)=r B(t,r)\label{B}
\end{eqnarray}
It must be noted that, for $C_{4}=1,\,C_{Z}=1,\,C_{1}=2C_{Z},\,\bar{b}(t)=\dot{f(t)}=-C_Z\,t,
\, \bar{a}=C_{3}/2,\, C(t)=-C_{2}$ these forms of the gravitational potentials obtained in \cite{ON}
\eqref{A}-\eqref{B} reduces to the obtained by the combination of this adhoc form of the anisotropy together with Karmarkar condition.
In the collapsing phenomena, the metric variables $a(r)$ and $b(r)$ should remains positive and nonzero. The condition $a(r)>0$ and $b(r)>0$ during the collapse implies that from equation \eqref{a(r)} we must have $C_{Z}^2<C_{4}^{2}\left[\frac{C_{1}}{C_{3}r^{2}+2C_{4}}-2C_{2}C_{3}\right]^2$.
\begin{figure}[!h]
\begin{subfigure}{.5\textwidth}
\centering
\includegraphics[width=\linewidth]{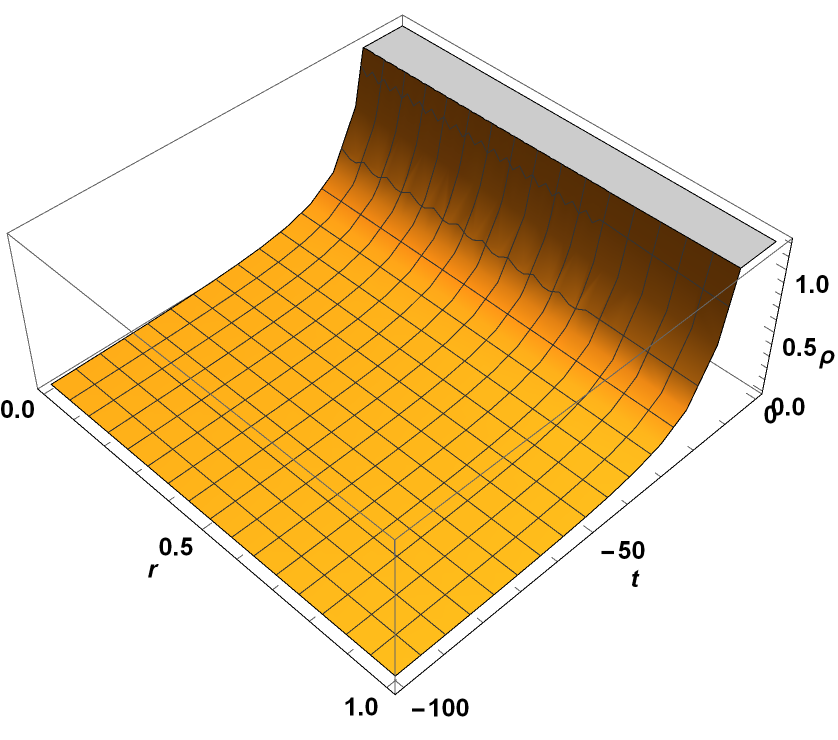}
\caption{}
\label{fig:rho}
\end{subfigure}
\begin{subfigure}{.5\textwidth}
\centering
\includegraphics[width=\linewidth]{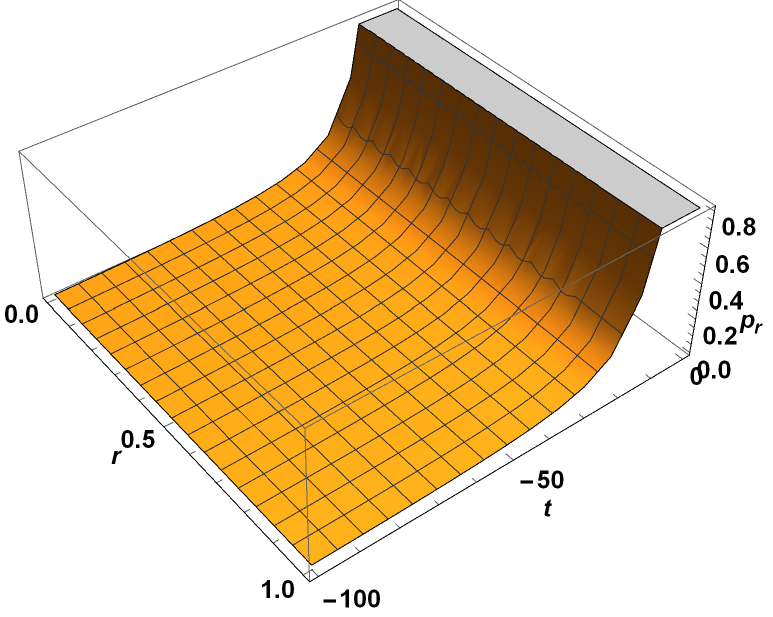}
\caption{}
\label{fig:pr}
\end{subfigure}
\caption{(a) $\&$ (b) Shows the plots of the density $\rho$ \eqref{rho} and radial pressure $p_{r}$ \eqref{pr}, with respect to $t$ and $r$, with center at $r=0$ and surface at $r=1$. It is observed that both density and radial pressure remains regular and positive throughout the collapse.}
\end{figure}

\begin{figure}[!h]
\begin{subfigure}{.5\textwidth}
\centering
\includegraphics[width=\linewidth]{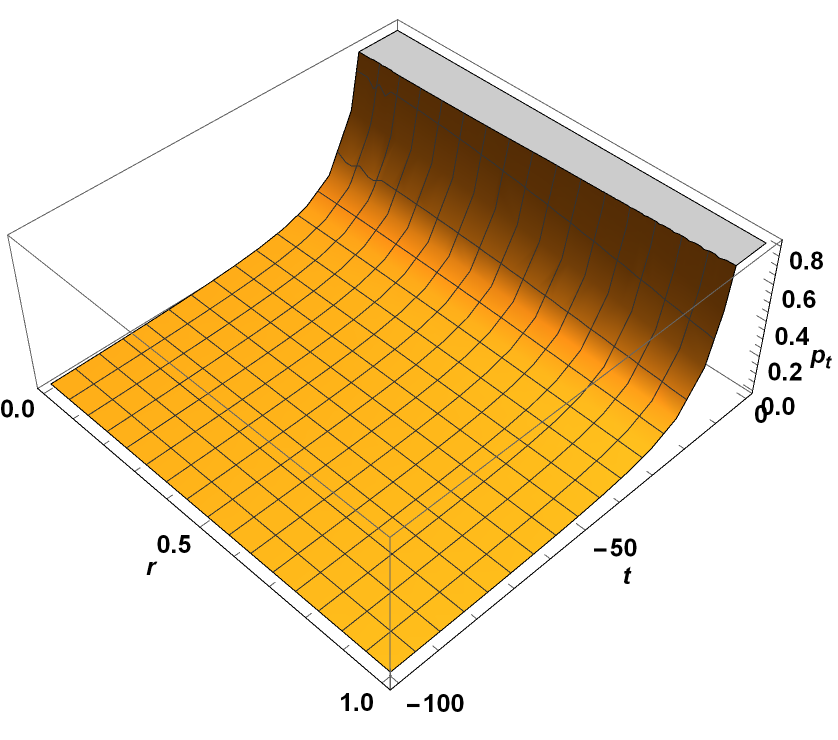}
\caption{}
\label{fig:pt}
\end{subfigure}\begin{subfigure}{0.5\textwidth}
\centering
\includegraphics[width=\linewidth]{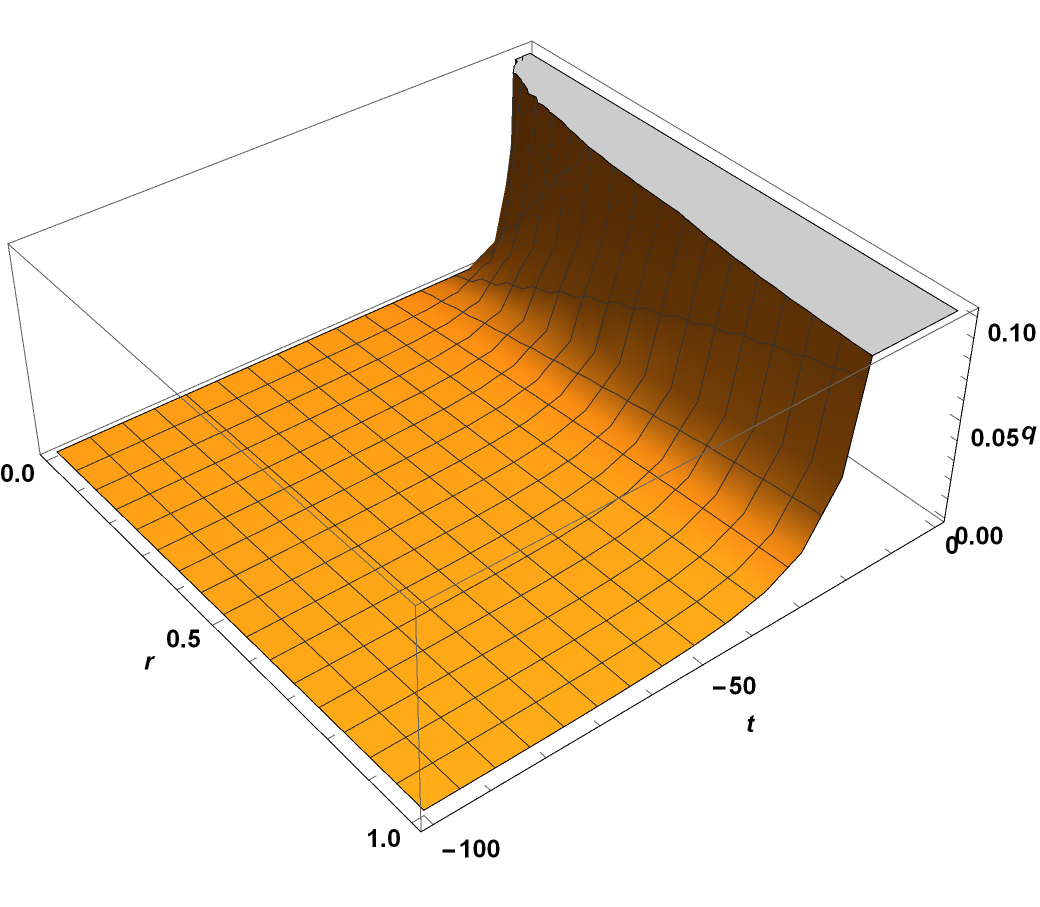}
\caption{}
\label{fig:q}
\end{subfigure}
\caption{(a) Plot of the tangential pressure $p_{t}$ \eqref{pt}, with respect to $t$ and $r$, with center at $r=0$ and surface at $r=1$. 
(b) Plot of the radial heat flux $q$ \eqref{q}, with respect to $t$ and $r$. 
It is observed that as the stellar system deviates from the equilibrium, it starts radiating more and more heat flux. 
Plots (a) $\&$ (b) shows that both tangential pressure and radial heat flux remains regular and positive throughout the collapse.}
\end{figure}
\begin{figure}[!h]
	\begin{subfigure}{.5\textwidth}
		\centering
		\includegraphics[width=\linewidth]{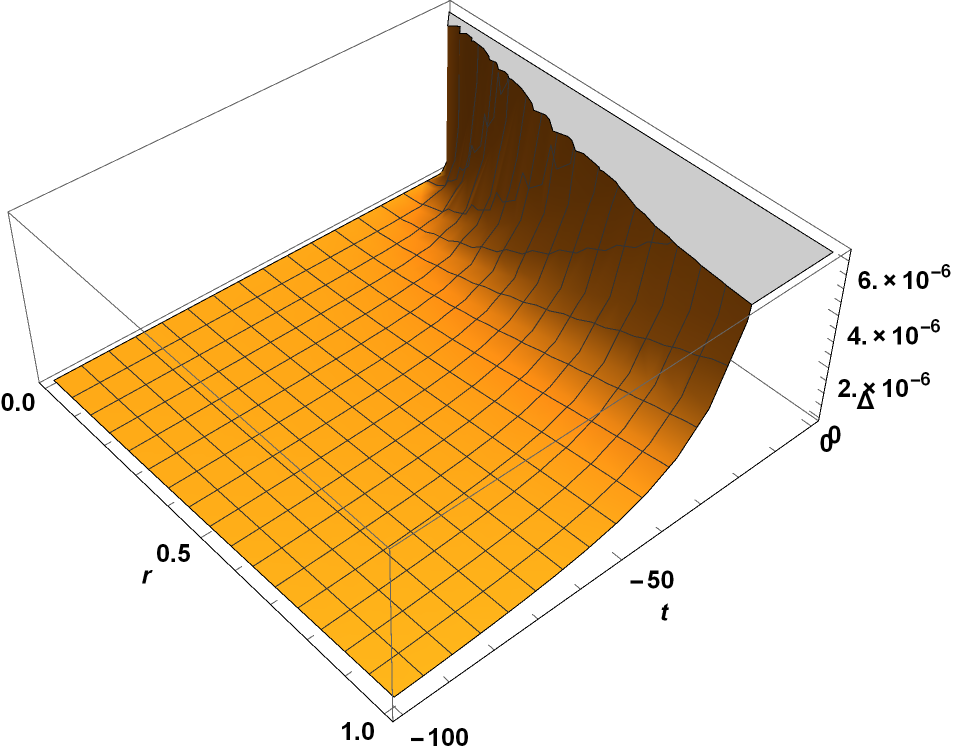}
		\caption{}
		\label{fig:delta11}
	\end{subfigure}\begin{subfigure}{0.5\textwidth}
		\centering
		\includegraphics[width=\linewidth]{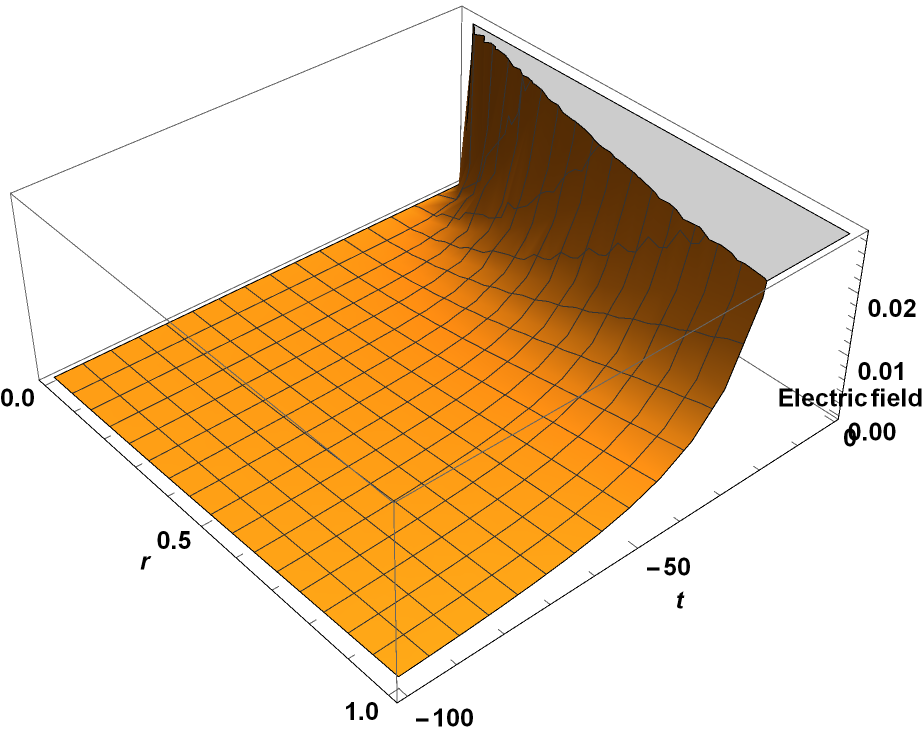}
		\caption{}
		\label{fig:EE}
	\end{subfigure}
	\caption{ (a) $\&$ (b) Shows the plots of the pressure anisotropy $\Delta$, equation \eqref{Delta1} and the Electric field with respect to $t$ and $r$. Both the pressure anisotropy and Electric field remains regular and positive throughout the collapse.}
\end{figure}
\begin{figure}[!h]
	\begin{subfigure}{.5\textwidth}
		\centering
		\includegraphics[width=\linewidth]{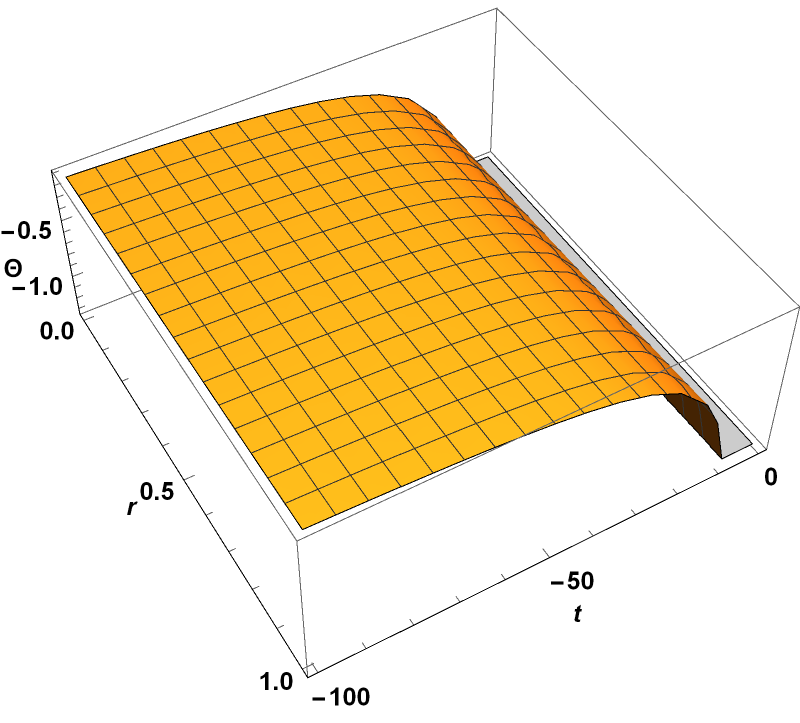}
		\caption{}
		\label{fig:Theta}
	\end{subfigure}\begin{subfigure}{0.5\textwidth}
		\centering
		\includegraphics[width=\linewidth]{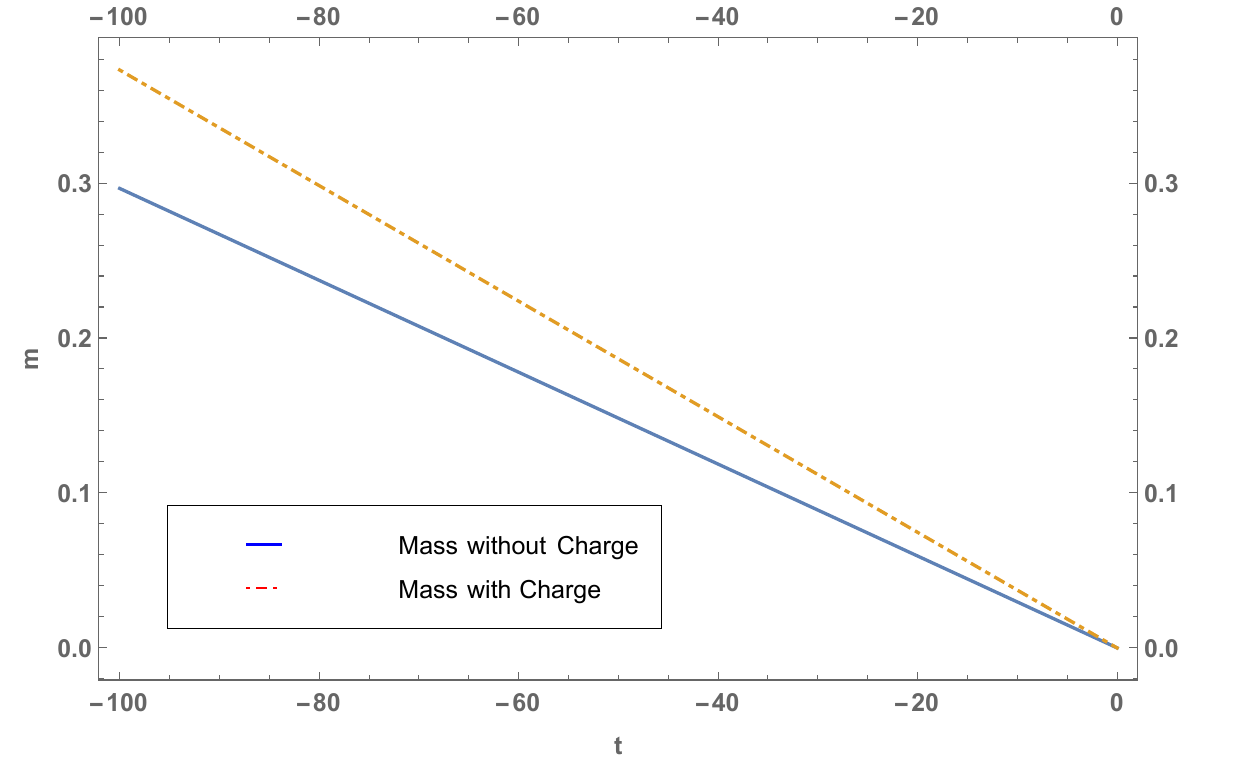}
		\caption{}
		\label{fig:m}
	\end{subfigure}
	\caption{ (a) Plot of the expansion scalar $\Theta$, equation \eqref{Thetaabf}, with respect to $t$ and $r$. As expected for the collapsing phenomena, the expansion scalar $\Theta$ remains negative throughout the interior of stellar system. 
		(b) Plot of the mass $m$, equation \eqref{mabf}, of the collapsing stellar system with respect to $t$ for both chargeless and charged collapse. It is observed that the mass radiates linearly for both the cases, however, it has large value for the charged case than that of its counterpart.}
\end{figure}

The expansion scalar \eqref{Theta} and the Misner sharp mass function \eqref{mass} have the form
\begin{eqnarray}
\Theta &=& \frac{6\sqrt{C_{3} C_{4}}}{t \sqrt{C_{4}^{2} \left(2 C_{2} C_{3}
		-\frac{C_{1}}{C_{3} r^{2}+2 C_{4}}\right)-2 C_{Z}^{2}}}\label{Thetaabf}\\
m&=&\frac{8 t r^{3} C_{3} C_{4}^3 C_{Z}}{\left(C_{3} r^{2}+2 C_{4}\right)^{3}}
\,\left[\frac{2 C_{2} C_{3} \left(C_{3} r^{2}+2 C_{4}\right)-C_{1}}
{2 \left(C_{3} r^{2}+2 C_{4}\right) \left(C_{2} C_{3} C_{4}^{2}-C_{Z}^{2}\right)-C_{1} C_{4}^{2}}\right]\label{mabf}
\end{eqnarray}
The boundary condition \eqref{matching} in the view of \eqref{pr}-\eqref{q} becomes
\begin{eqnarray}
2f\,\ddot{f}+\dot{f}^2-2x\dot{f}&=&y \label{match}
\end{eqnarray}
where 
\begin{eqnarray}
x&=&\left(\frac{a^{'}}{b} \right)_{\Sigma}\label{x}\\
y&=&\left(\frac{a^2}{b^2}\left[\frac{b^{'\,^{2}}}{b^2}+\frac{2}{r}\left(\frac{b^{'}}{b}
+\frac{a^{'}}{a}\right)+\frac{2a^{'}b^{'}}{ab} \right] \right)_{\Sigma} \label{y}
\end{eqnarray}
\newpage
\section{Stability and Energy conditions}\label{sec4}
To check the stability of the shear free condition for our model, we consider the stability criteria given in \cite{SC1964, Herrera:1992lwz, Abreu:2007ew}. Here, from the study of the behavior of adiabatic index and speed of sound one can infer the stability of the stellar system. 
More precisely, a stellar model is considered to be stable (unstable) if adiabatic index $\Gamma$ is greater than 4/3 ($\Gamma<4/3$). In this range the pressure remains greater (less) than that of the weight of stellar system and hence, it remains in equilibrium, any departure from equilibrium leads to collapse. Other way to study the stability of a stellar system is by using the difference between the propagation of the radial $V_{r}^{2}$ and transverse $V_{t}^{2}$ speeds of sound in a stellar system. Both the speeds of sound should always be less than that of velocity of light and lies in the range $0<V_{r}^{2}$ and $\,V_{t}^{2}<1$.
Also, as shown in \cite{Herrera:1992lwz, Abreu:2007ew}, we can check whether a particular region of a matter configuration is potentially stable or unstable. The regions where the radial speed of sound remains greater than that of the transverse speed are called as the potentially unstable regions and viceversa \cite{Abreu:2007ew}.

\begin{figure}[!ht]
\begin{subfigure}{0.3\textwidth}
	\centering
	\includegraphics[width=\linewidth]{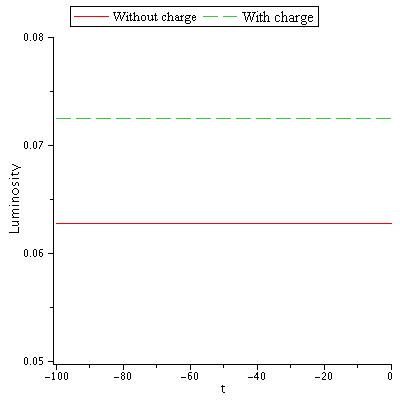}
	\caption{}
	\label{fig:Luminosity}
\end{subfigure}
\begin{subfigure}{0.3\textwidth}
	\centering
	\includegraphics[width=\linewidth]{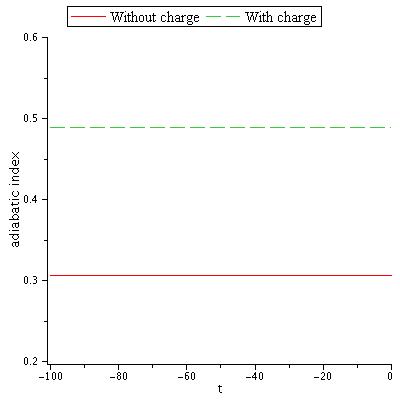}
	\caption{}
	\label{fig:Adiabatic}
\end{subfigure}
\begin{subfigure}{0.3\textwidth}
	\centering
	\includegraphics[width=\linewidth]{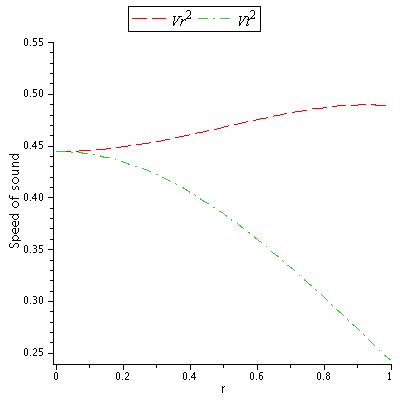}
	\caption{}
	\label{fig:Speedofsound}
\end{subfigure}
	\caption{ (a) $\&$ (b) shows the plots of the luminosity and adiabatic index with respect to $t$ for both the chargeless and charged collapse. It has been found that both luminosity and adiabatic index are time independent. The adiabatic index $\Gamma$ has value less than 4/3 implying that the stellar system under consideration is representing unstable and collapsing phenomena. (c) shows the plots of the speed of sound and can be seen that both radial $V_{r}^{2}$ and tangential $V_{t}^{2}$ speed of sound lies in the range of the $0<V_{r}^{2}\,; \,V_{t}^{2}<1$. The stellar system is unstable as the radial speed remains greater than that of the tangential speed. Thus, the unstable and collapsing nature of the stellar system is confirmed by the behavior of adiabatic index Fig. (b) and speed of sound Fig (c).}
\end{figure}

It can be seen from adiabatic index $\gamma$, Fig \ref{fig:Adiabatic} that it remains a constant and less than $4/3$ implying the stellar system under consideration is unstable and represents a collapsing phenomena. The unstable and collapsing nature of the stellar system is also confirmed from the behavior of radial $V_{r}^{2}$ and transverse $V_{t}^{2}$ speeds of sound. Throughout the interior of the stellar system, the radial speed of sound remains greater than that of transverse speed and it shows that the region of matter configuration is potentially unstable.
It can be seen from the Figs. \ref{fig:rho}, \ref{fig:pr} and \ref{fig:pt}
that density, radial pressure and tangential pressure remains regular and positive throughout the charged collapse. Also, radial heat flux is regular throughout the collapse as can be seen from Fig. \ref{fig:q}. 
It can be seen that pressure anisotropy \ref{fig:delta11} and electric field intensity \ref{fig:EE} remains regular and positive.
The expansion scalar $\Theta$ shows the contracting behavior as show in Fig. \ref{fig:Theta} it remains regular and negative throughout the collapse.
Fig. \ref{fig:m} shows mass $m$, equation \eqref{mabf}, of the collapsing stellar system with respect to $t$ for both chargeless and charged collapse. It is observed that the mass radiates linearly for both the cases, however, it has large value for the charged case than that of its counterpart.
Similar to our earlier work of noncharged case \cite{Jaryal:2020mag}, the Luminosity \eqref{LInfinity} of the charged radiating collapse is time independent.
Figs. \ref{fig:Adiabatic} and \ref{fig:Speedofsound} shows that the stellar system to be unstable and representing collapsing phenomena.

\begin{figure}[h]
\begin{subfigure}{0.3\textwidth}
\centering
\includegraphics[width=\linewidth]{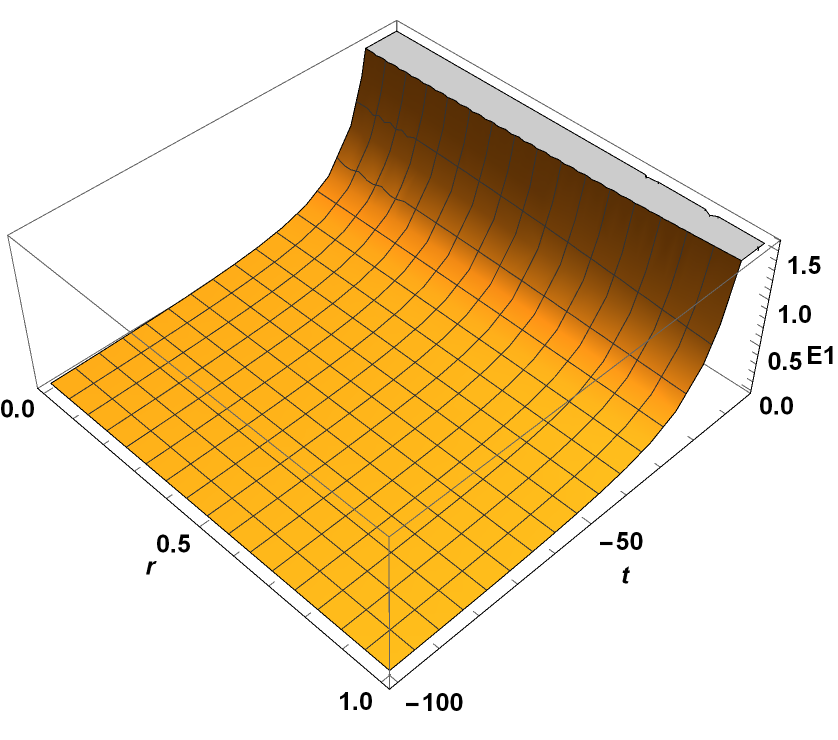}
\caption{}
\label{fig:E1}
\end{subfigure}
\begin{subfigure}{0.3\textwidth}
\centering
\includegraphics[width=\linewidth]{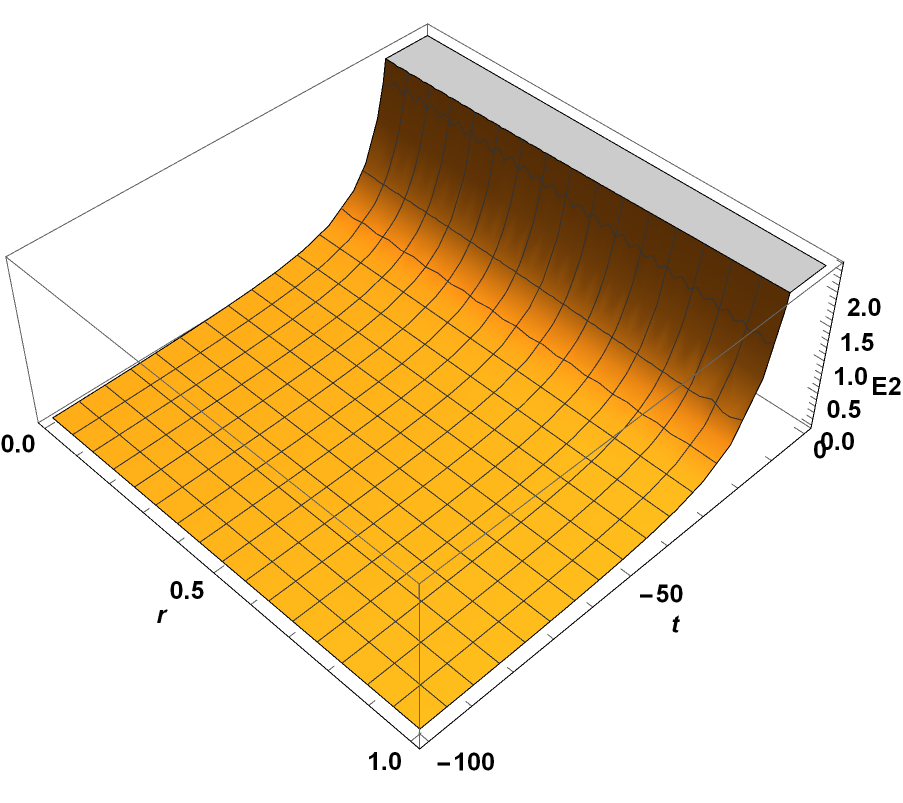}
\caption{}
\label{fig:E2}
\end{subfigure}
\begin{subfigure}{0.3\textwidth}
\centering
\includegraphics[width=\linewidth]{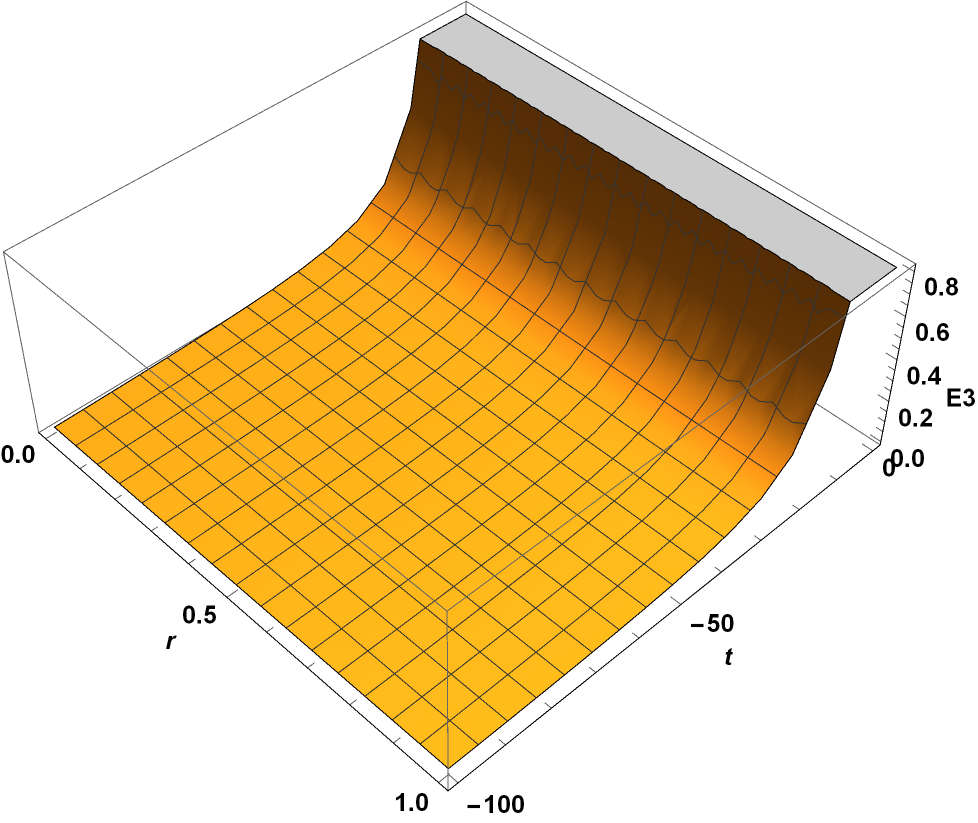}
\caption{}
\label{fig:E3}
\end{subfigure}
\caption{ (a), (b) $\&$ (c) shows the plots of energy condition $E1,\, E2, \, E3$ for the charged radiating star with respect to $t$ and $r$ respectively. It shows that this stellar system satisfies these above conditions.}
\end{figure}

Let us now ask if the solutions also satisfy the energy conditions as well.
Energy conditions plays important role in the study of the astronomical phenomena 
like collapsing stellar models. In this section we will analyze the physical
evidences of our model by verifying the energy conditions. The energy conditions
namely weak energy condition\,(WEC), null energy condition\,(NEC), 
dominant energy condition\,(DEC) and strong energy conditions \,(SEC) 
will be satisfied at all points in the stellar model if the following
inequalities are satisfied simultaneously~\cite{Chan1997}\\
{\bf{E1\,:}}\, $\left(\rho+p_{r}\right)^2-4q^{2}\,\geq \,0$\hspace{3.2cm}(SEC/DEC/WEC)\\
{\bf{E2\,:}}\, $\rho-p_{r}\,\geq\,0\,\,\,\, $\hspace{4.3cm} (DEC)\\
{\bf{E3\,:}}\, $\rho-p_{r}-2p_{t}+\sqrt{\left(\rho+p_{r}\right)^2-4q^{2}}\,\geq\,0 $\hspace{.2cm} (DEC)\\
{\bf{E4\,:}}\, $\rho-p_{r}+\sqrt{\left(\rho+p_{r}\right)^2-4q^{2}}\,\geq\,0\,\,\,\, $\hspace{.9cm} (DEC/WEC)\\
{\bf{E5\,:}}\, $\rho-p_{r}+2p_{t}+\sqrt{\left(\rho+p_{r}\right)^2-4q^{2}}\,\geq\,0$\hspace{.2cm} (SEC/DEC/WEC)\\
{\bf{E6\,:}}\, $2p_{t}+\sqrt{\left(\rho+p_{r}\right)^2-4q^{2}}\,\geq\,0$\hspace{1.7cm} (SEC)\\
Beside these energy conditions, a physically reasonable stellar model should also satisfy \\
{\bf{E7\,:}}\, $\rho>0$,\, $p_{r}>0$,\, $p_{t}>0$, and $\rho^{'}<0$,\, $p_r^{'}<0$,\, $p_{t}^{'}<0$.\\
Here, we can see that the validity of the {\bf{E1}} and {\bf{E2}} inequalities implies that
the {\bf{E4}} inequality is satisfied. In the same fashion, the validity of the {\bf{E1}},
{\bf{E2}} and {\bf{E7}} inequalities ensures that the {\bf{E5}} and {\bf{E6}} inequalities
are satisfied. So, in general we only need to check the validity of the 
{\bf{E1}},\, {\bf{E2}},\,{\bf{E3}} and {\bf{E7}}. For our radiating stellar model,
it can be seen from the figures \ref{fig:E1}, \ref{fig:E2} and \ref{fig:E3} 
that all these energy conditions are well satisfied throughout the interior of the collapsing star.
\section{Thermal properties }\label{sec5}
The study of the thermodynamical evolution and the temperature
profiles of the radiating stars plays prominent role
during dissipative gravitational collapse as they decide the departure from the
thermodynamical equilibrium. Previous studies of the shear free and
shearing dissipative gravitational collapses shows that the relaxation 
effects plays significant role in the temperature profiles towards the end state of the 
dissipative gravitational collapse
see \cite{Maartens1995,Martinez1996, Herrera1996, Herrera1997, LHerrera, GMM1998, GMM1999} and references therein.
To study the thermodynamical evolution and the temperature profiles 
of inside the collapsing star, we will use the causal transport equation for the metric 
\eqref{1eq1}  given by\,\cite{Maartens1995, Martinez1996, LHerrera, GMM1998}
\begin{eqnarray}
\tau h_{\mu}^{\nu}\dot{q}_{\nu}+q_{\mu}&=&-k\left( h_{\mu}^{\nu} \nabla_{\nu} T+T\dot{u}_{\mu} \right)\label{tempgen}\\
\tau \left(qbf\right)_{,t}+q\,a\,b\,f&=&-\frac{k\left(aT \right)_{,r}}{bf} \label{tempabf}
\end{eqnarray}
where, $\alpha>0$,\,$\beta>0$,\,
$\gamma>0$ and $\sigma>0$ are constants and $h^{\mu\nu}=g^{\mu\nu}+u^{\mu}u^{\nu}$.
Also,
\begin{eqnarray}
 \tau_{c}=\left(\frac{\alpha}{\gamma}\right)\,T^{-\sigma}\hspace{0.2cm},\hspace{0.6cm} 
 k=\gamma\, T^{3}\,\tau_{c}\hspace{0.2cm},\hspace{0.6cm}  \tau=\left(\frac{\beta\,\gamma}{\alpha}\right)\,\tau_{c} \label{tkt}
\end{eqnarray}
 are physically reasonable choices of the mean collision time, between 
massive and massless particles $\tau_{c}$\,,\, thermal conductivity $k$ and 
the relaxation time $\tau$ respectively\,\cite{Martinez1996, GG2001}. 
$\tau$ represents the causality index, measures the strength of relaxational effects
and $\tau=0$ or $\beta=0$ represents the noncausal case.
Using these forms of the physical quantities given in equation \eqref{tkt},
the form of the causal heat transport equation \eqref{tempabf} becomes
\begin{eqnarray}
\beta T^{-\sigma} \left(qbf\right)_{,t}+q\,a\,b\,f&=&-\frac{\alpha \,\left(aT \right)_{,r}}{bf}\,\,T^{3-\sigma} \label{tempabff}
\end{eqnarray}
The noncausal solution of heat equations are obtained by setting
$\beta=0$ i.e. $\tau=0$ in the above transport equation \eqref{tempabff}\,\cite{GG2001}
\begin{eqnarray}
\left(a\,T\right)^{4}&=&-\frac{4}{\alpha}\int a^{4}\,q\,b^2\,f^2\,dr+F(t),
\hspace{1.6cm} \sigma= 0\label{Nsigman4}\\
\ln\left(a\,T\right)&=&-\frac{1}{\alpha}\int q\,b^2\,f^2\,dr+F(t),\hspace{2cm} \sigma=4\label{Nsigma4}
\end{eqnarray}

The causal solution of the above transport equation \eqref{tempabff} are given by\,\cite{GG2001}
\begin{eqnarray}
\left(a\,T\right)^{4}&=&-\frac{4}{\alpha}\left[\beta\int a^3\,b\,f(q\,b\,f)_{,t}\,dr
+\int a^{4}\,q\,b^2\,f^2\,dr\right]+F(t), \hspace{0.6cm} \sigma=0\label{Csigman4}\\
\left(a\,T\right)^{4}&=&-\frac{4\beta}{\alpha}exp\left(-\int\frac{4\,q\,b^{2}\,f^{2}}{\alpha}\,dr\right)
\int a^3\,b\,f(q\,b\,f)_{,t}\,dr\,exp\left(\int\frac{4\,q\,b^{2}\,f^{2}}{\alpha}\,dr\right)\nonumber\\
&&+F(t)exp\left(-\int\frac{4\,q\,b^{2}\,f^{2}}{\alpha}\,dr\right),\hspace{4.2cm} \sigma=4\label{Csigma4}
\end{eqnarray}
where $F(t)$ is the function of integration. The function $F(t)$ is determined by invoking boundary conditions
\begin{eqnarray}
\left(T^{4} \right)_{\Sigma}&=&\left( \frac{L_{\infty}}{4\pi\delta r^{2}b^{2}f^{2}} \right)_{\Sigma} \label{Tsigma}
\end{eqnarray}
where $L_{\infty}$ is the total luminosity for an observer at infinity and $\delta>0$ is constant. 
\begin{figure}[h]
	\begin{subfigure}{.5\textwidth}
		\centering
		\includegraphics[width=\linewidth]{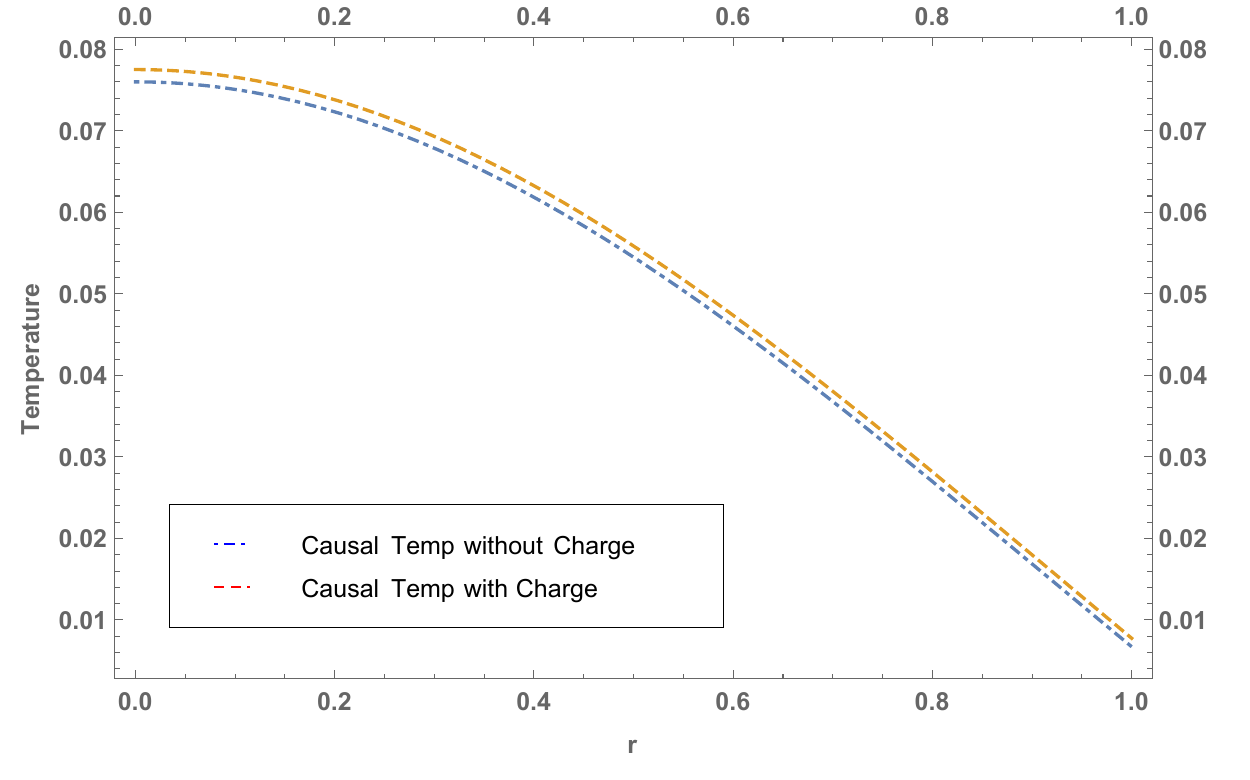}
		\caption{}
		\label{fig:Temp}
	\end{subfigure}\begin{subfigure}{0.5\textwidth}
		\centering
		\includegraphics[width=\linewidth]{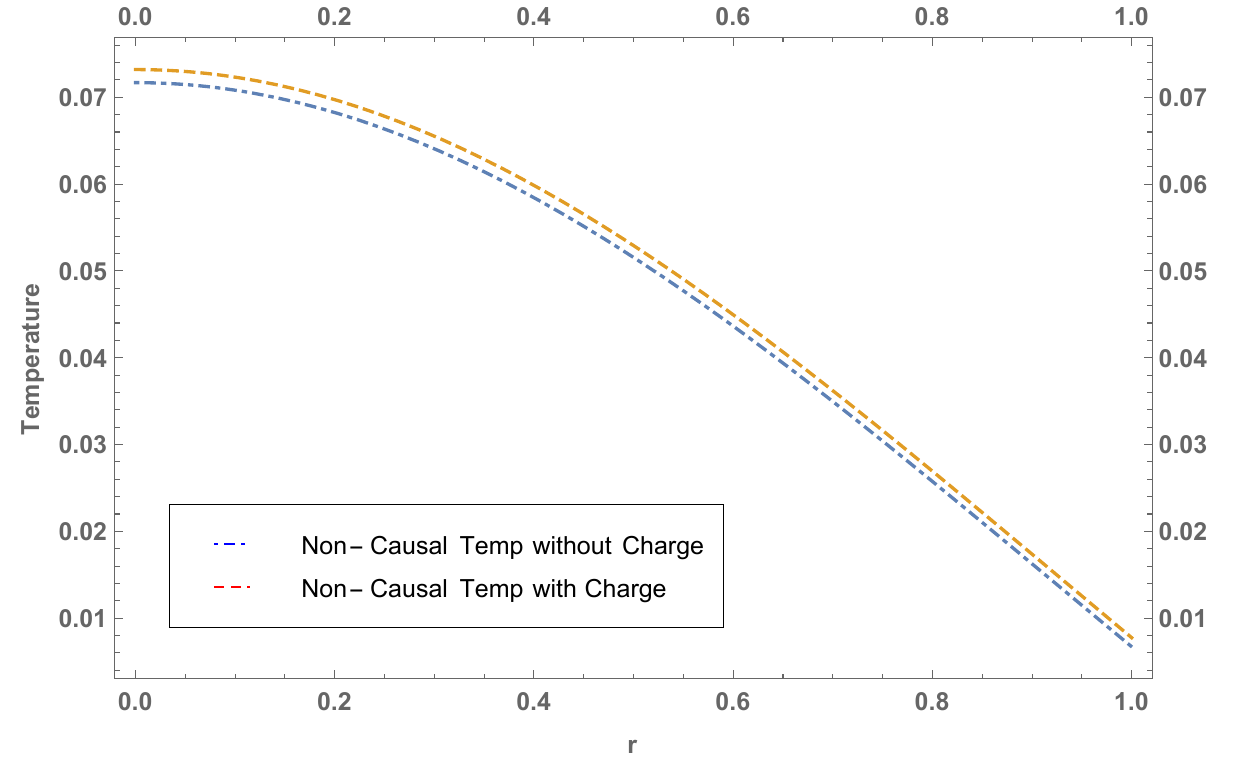}
		\caption{}
		\label{fig:Tempf2}
	\end{subfigure}
	\caption{(a) $\&$ (b) shows the plots of causal and non- causal temperatures $T$ of stellar system with respect to $t$ and $r$. It is observed that both causal and non- causal temperatures for the charged collapse remains greater than that of their counterpart, chargeless collapse. It is also observed that the causal temperature remains greater than that of non- causal temperature throughout the interiors of both the charged and chargeless collapse.
	}
\end{figure}
As with previous investigations\,\cite{GMM1998, GG2001}, Fig.\ref{fig:Temp} 
shows that both the causal and noncausal temperature are same at the boundary of the star.
However, at later stages of the collapse, relaxation effects plays significant role 
and they differ from the noncausal case. This behavior can be seen from the 
Fig. \ref{fig:Temp} that with $\beta>0$ the 
relaxations effects grows and the causal temperature remains greater than that 
of noncausal temperature throughout the interior of the star.
These results are in agreement with the earlier results obtained for 
the shear free collapse\,\cite{LHerrera, GMM1999}. 
\section{Conclusion}\label{sec6}
In this paper we have studied the shear free spherical symmetric gravitational collapse of charged radiating astrophysical systems including compact configurations on the stellar or galactic scales. The interior and exterior spacetimes of the solution has been smoothly matched over the timelike hypersurface. To exactly solve for the system,
we have utilised the pressure anisotropy \eqref{Delta1} together with the Karmarkar condition \cite{Karmarkar}, where the metric functions are taken to be linear in time dependence. 
 
Let us describe our observations and conclusions obtained here: First, the model under consideration has 
all the characteristics of a physically viable compact object, since all the physical quantities 
like density \eqref{rho} pressure profiles \eqref{pr}, \eqref{pt}, radial heat flux \eqref{q}, pressure anisotropy \eqref{Delta1} and Electric filed intensity are regular and positive throughout the collapse. The mass \eqref{mabf} is a linear function of time. It is observed that the mass for the charged collapse is greater than that of the chargeless case. One may also show that the luminosity of this system is time independent and radiates uniformly throughout the collapse. Secondly, we confirm that the system is truly unstable. To verify, we determined the stability of the charged radiating stellar system by studying the behavior of the adiabatic index. This value equals the difference of the radial and transverse speeds of sound in the matter configuration. It is known that a matter configuration is stable (unstable) if the adiabatic index is greater than $4/3$ ($\Gamma<4/3$) \cite{SC1964, Herrera:1992lwz} . For the present system, the adiabatic index for the stellar system under consideration is less $4/3$, implying that the system is unstable and represents a collapsing phenomena. Thirdly, as can be seen from the Figs \ref{fig:E1},\,\ref{fig:E2} and\,\ref{fig:E3} that the model under consideration satisfies the energy conditions for the charge radiating stellar system. This also strengthens our claim that the model is physically viable.
 Furthermore, the thermal properties of the system show that the causal and non- causal temperature for the charged collapse are greater than that of their counterpart of chargeless collapse; the causal temperature remains higher than that of the non- causal temperature throughout the collapse.

The presence of charge has great importance in the study of radiating stellar system. It affects significantly the behavior of all the physical quantities like density, pressure and mass profiles during the collapsing phenomena. The stability analysis also gets affected by the presence of charge in the collapse, as adiabatic index and radial $\&$ transverse speeds of sound has significant difference to their counterpart chargeless case. The study of thermal properties shows that both causal and non- causal temperatures for charge collapse has higher value than that of chargeless collapse. Thus collapse with charge does affects the collapsing phenomena as it delays the collapse and it takes more time to reach the central singularity than that of the chargeless collapse.https://www.overleaf.com/project/64f19e779bcc28ba58f98a68
\\\\{\bf{Acknowledgments}}\\
The author AC  is supported through the DAE-BRNS project 58/14/25/2019-BRNS.

\end{document}